# Passive thermal fine-tuning of whispering gallery mode for nonlinear optics in a microcavity

Yaming Feng,[1)] Yuanlin Zheng,[2,3)] Fangxing Zhang,[2)] Jianfan Yang,[2)] Tian Qin,[2)] and Wenjie Wan[1,2)]

[1]*The State Key Laboratory of Advanced Optical Communication Systems and Networks and Collaborative Innovation Center of IFSA, School of Physics and Astronomy, Shanghai Jiao Tong University, Shanghai 200240, China*

[2] *MOE Key Laboratory for Laser Plasmas, the University of Michigan-Shanghai Jiao Tong University Joint Institute, Shanghai Jiao Tong University, Shanghai 200240, China*

[3]*Department of Electrical and Systems Engineering, Washington University in St. Louis, St. Louis, Missouri 63130, USA.*

**Whispering-gallery-mode (WGM) microcavities strongly enhance nonlinear optical processes like optical frequency comb, Raman scattering and optomechanics, which nowadays enable cutting-edge applications in microwave synthesis, optical sensing spectroscopy, and integrated photonics. Yet, tunability of their resonances, mostly via coarse and complicated mechanism through temperature, electrical or mechanical means, still poses a major challenge for precision applications as above. Here we introduce a new passive scheme to finely tune resonances of WGMs at MHz precision with an external probe. Such probe remotely transfers heat through a gap from an optical microcavity, effectively tuning its resonances by thermal-optic nonlinearity. Moreover, we explore this unique technique in microcavity nonlinear optics, demonstrating the generation of a tunable optical frequency comb and backward stimulated Brillouin scattering with variable beating frequencies. This new technique addresses the core problem of WGM microcavity's fine-tuning, paving the way for important applications like spectroscopy and frequency synthesis.**

Whispering-gallery-mode (WGM) micro-cavities have been intensively investigated in a wide range of areas for both fundamental physics and practical applications. The ultra-high quality (Q) factors and ultra-small mode volumes make WGM micro-cavities ideal platforms for ultra-sensitive optical sensing, strongly enhanced nonlinear optics, cavity QED, optomechanics, and micro-scale optical frequency comb [1-5]. However, one major obstacle halting the practical implementations in these applications is the lack of tunability for fixed microstructures unlike their bigger-scale opponents, i.e. Fabry–Pérot cavities. For example, WGMs have to be finely tuned to closely match with atomic lines in cavity QED [6]; dual-comb spectroscopy requires precision scanning of optical comb lines [7]; tunable microwave generation from SBS enabled sources also poses a great demand for tunability [8]. Previously, various approaches have been demonstrated to tune WGM resonances mainly through thermal, mechanical, electrical and optical means [9-12]. Active thermal tuning through thermo-optic effect can be easily implemented for a wide tuning range. For example, WGM shift of 160 pm can be achieved in a temperature change of 10 K for a silica microsphere [13]. Mechanically, compressing or stretching of micro-cavities by external piezo-electric components has been widely reported to even cover a whole free spectral range (FSR) tuning [10,14]. Besides, On-chip carrier injection method for tuning and modulation, however, only works for specific semiconductor like silicon [15]. However, integration of these active tuning elements remains difficult.

Among the above methods, thermo-optic tuning of WGMs is one of the popular methods for the resonance tuning. Thermo-optic tuning based on electrical heating has exhibited a tuning rate excess $85 \text{ GHz/V}^2$ over a 300 GHz range for microtoroid [16]; similarly, silicon photonic micro-disks can also be tuned by heating the graphene covered surface [17], or just through the heat-pad nearby [18]; laser or optical induced heating can also be applied to thermal tuning, either through microcavities' intrinsic material absorption, e.g. by a CO2 laser [19], or by attaching special light absorbers, nanoparticles to the surfaces [20,21] during light illumination. However, all the above methods have to rely on active external sources, i.e. electric, light or heat and complex structures, which makes them inconvenient to implement.

Here, we demonstrate a passive mechanism to achieve extremely fine-tuning at MHz level of WGM resonances through thermal optical effect only using an external probe. The probe servers as a heat sink to dissipate absorbed heat from an optical microcavity, tuning its resonances by altering temperature inside. Such heat transfer based tuning scheme highly depends on the gap spacing as well as probes' materials. We can achieve up to a 300 MHz spanning range at MHz resolution by fine positioning a silica fiber probe within 500 μm distance. We also found out a better tuning efficiency by a metallic cupper probe with a larger thermal conductance. Moreover, by applying this new technique, we show that a tunable optical frequency comb with a variable family comb lines can be generated, which is hard to achieve previously by laser pumping or detuning. In addition, we demonstrate a tunable microwave frequency source based on cavity-enabled backward stimulated Brillouin scattering by this new tuning scheme. Unlike previous active methods, the current precise method is passive and heat-source free, it should be generally applicable for microcavity based applications, opening a new avenue for important applications like spectroscopy and frequency synthesis.

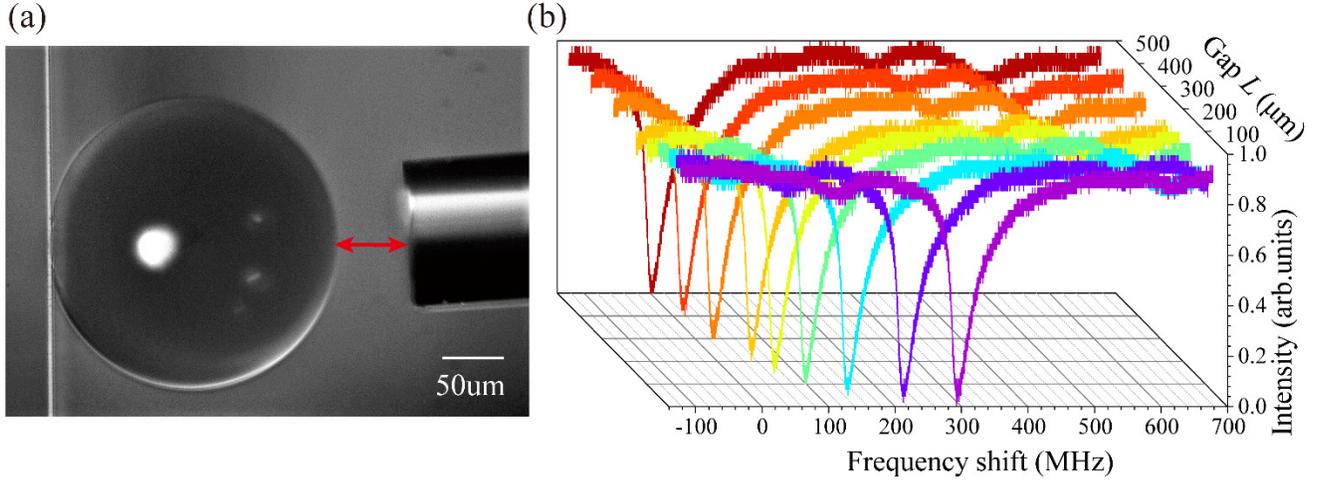

FIG. 1. The thermal fine-tuning apparatus and the related results. (a) Top view of the microcavity system. The gap distance $L$ between the microsphere and the fiber probe is adjustable. (b) The WGM resonance frequency shift varies as changing the gap distance $L$. Up to 300MHz shift can be realized with adjusting the gap in a range of 500 μm.

In a common thermal tuning WGM scheme, the thermal effect induces the temperature-dependent resonance frequency shift $\Delta f$ from the refractive index n and the diameter D variations in a fused silica microsphere, as shown below [22]:

$$\Delta f = -f_0 \left( \frac{1}{n}\frac{dn}{dT} + \frac{1}{D}\frac{dD}{dT} \right) \Delta T, \qquad (1)$$

where $f_0 = c/\lambda_0$ is the cold microcavity resonant frequency at $T_0 = 293.15$ K, and $n = 1.46$ is the refractive index of fused silica for $\lambda_0 = 1550$ nm. $dn/ndT = 8.52 \times 10^{-6}$ K$^{-1}$[23] and $dD/DdT = 0.55 \times 10^{-6}$ K$^{-1}$ [24] are thermo-optic coefficient and thermal expansion coefficient, respectively, which indicates the major contribution is from the thermo-optic effect. For a silica microsphere, the temperature change $\Delta T$ is caused by the absorption of the resonant optical power coupled into the microcavity, such that the input heat $Q_{in}$ will equal the outgoing heat flow, i.e. $K_{eff}\Delta T$ at thermal equilibrium:

$$Q_{in} = K_{eff}\Delta T, \qquad (2)$$

where $K_{eff}$ [J/(s·K)] is the effective absolute thermal conductivity contributed from the surrounding air, microsphere's structure as well as supporting pillar [25,26]. From this perspective, if we purposely alter a microcavity's $K_{eff}$ by modifying its surround medium, structures or even supporting pillar, we may achieve a passive method without additional heating sources to change the internal temperature of the microcavity, effectively tuning the resonances.

Figure 1 shows one possible way to tune such microcavity resonances by placing an external solid probe in the vicinity. $K_{eff}$ then becomes adjustable when varying the gap distance $L$ between the probe and the microcavity. As a result, the WGM resonances will shift according to the internal temperature detuning caused by $K_{eff}$ changing at thermal equilibrium. Here an ultrahigh Q ($\sim 10^7$) silica microsphere with a diameter of around $240\ \mu m$ is evanescently coupled by a tapered fiber. In the meantime, a standard optical silica fiber probe ($125\ \mu m$ in diameter) is placed near the microsphere to induce the thermal disturbance. The gap distance $L$ between the micro-cavity and the fiber tip is varied form $5\ \mu m$ to $500\ \mu m$ by a translation stage with a micro-meter resolution. A pump laser of $1550\ nm$ wavelength (New Focus, TLB-6700) is swept around one WGM resonance with a frequency span of $2\ GHz$ at a rate of $20\ Hz$. Fig. 1(b) shows a typical resonance frequency shift with respect to gap variation by tuning the fiber tip. We observe up to repeatable $292\ MHz$ tuning range with an input power of $90\ \mu W$ during the gap variation from $5\ \mu m$ to $500\ \mu m$. Here we achieve one easy and passive way to fine tune WGM resonances at MHz resolution. Finer precision tuning, i.e. sub-MHz can be obtained if the probe is positioned in a large distance using a nanometer-stage to control the gap distance.

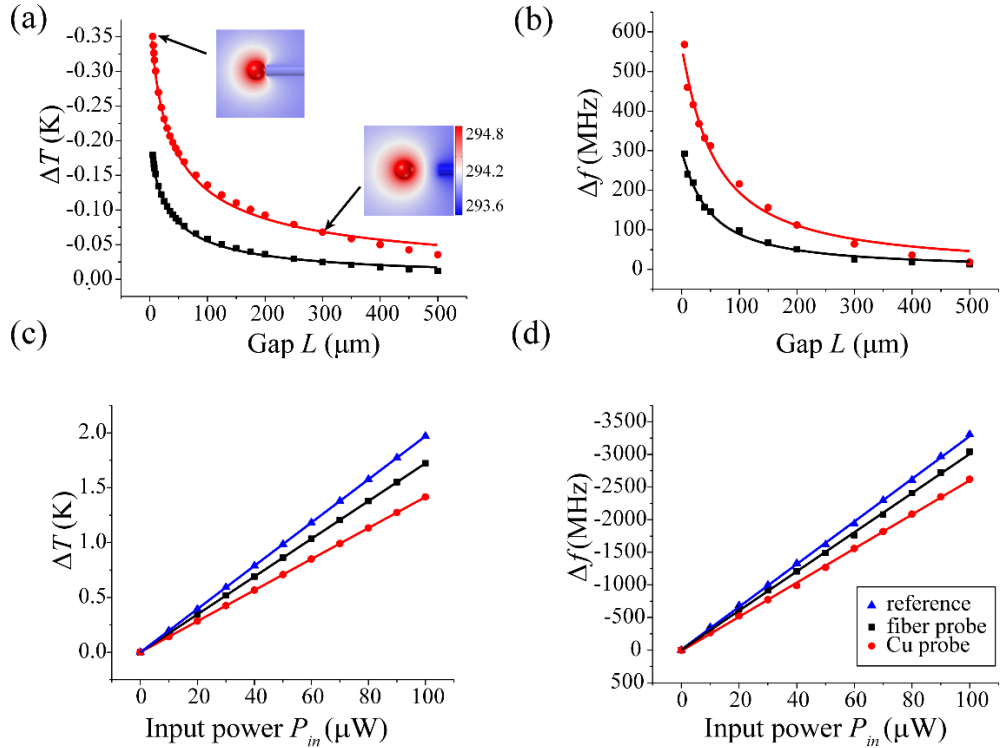

FIG. 2. The numerical simulations of temperature changes and the corresponding to frequency shift measured by the experiment. (a) The temperature change of microcavity $\Delta T$ varies with adjusting gap distance $L$. The insets are the temperature distribution of the microcavity system for Cu probe with $L = 5\ \mu m$ and $300\ \mu m$, respectively. (b) The frequency shift $\Delta f$ varies with gap $L$. The maximum values of $\Delta f$ for Cu probe is $568.10\ MHz$ larger than the one of fiber probe $292.26\ MHz$ at $L = 5\ \mu m$, even the diameter of Cu probe $D_{cu} = 100\ \mu m$ is smaller than the one of fiber probe $D_{fiber} = 125\ \mu m$. (c) The temperature shift $\Delta T$ of microcavity increases due to enlarging the input

power $P_{in}$. The Cu probe is much effective to reduce $\Delta T$ than the fiber tip. The case of fiber probe at $L = 800$ μm is also shown as a reference. The rates of the three lines (from up to down) are 19.69 mK/μW, 17.22 mK/μW and 14.15 mK/μW, respectively. (d) The experimental data of frequency shift $\Delta f$ changing with $P_{in}$. The rates of the three carves are $-33.04$ MHz/μW, $-30.40$ MHz/μW and $-26.16$ MHz/μW, respectively. The blue triangle, black square and red circle lines present: reference fiber, fiber and Cu probes, respectively. The solid lines are the fitting curves.

Figure 2 exams the cause of such fine-tuning by considering heat transfer between the cavity and the probe. To model such thermal tuning effect through gap varying, we perform finite element method (FEM) based heat transfer numerical simulations as comparison to experimental results [27]. Here the incoming heat flow $Q_{in}$ results from the input optical power $P_{in}$ through absorption with an effective mode area $A_{eff} = 78.5$ μm$^2$ (calculated from a WGM optical mode). The outgoing heat flows out of the cavity through the supporting pillar, the surrounded air as well as the external probe nearby. The surrounding air has a specific thermal conductance of $K_{air} = 0.026$ W/(m·K), far less than the materials we choose as the probe such as silica $K_{sio2} = 1.38$ W/(m·K) and cupper $K_{cu} = 400$ W/(m·K). Hence the external probe serves as a good heat sink to transfer and dissipate the heat from the cavity. The gap spacing between them becomes very crucial to determine the whole system's effective absolute thermal conductivity $K_{eff}(L)$, which is hard to estimate analytically. Instead, for the given structure, we perform a FEM based numerical method to calculate the temperature change $\Delta T$ with respect to the gap varying, then compare them to our experiments through $\Delta f/\Delta T = -1.756$ GHz/K calculated by Eq. (1) in our experiment. In Fig.2(a), for a silica fiber probe, $\Delta T$ varies almost linearly with a rate 0.1 mK/μm when $L > 200$ μm, then it behaves approximately as $1/L$, reaching to its maximum $\Delta T_{fiber-max} = -0.18$ K at $L = 5$ μm. For a Cu probe, a similar trend can be observed while with a larger amplitude: the maximum $\Delta T_{Cu-max}$ is $-0.35$ K achieved at $L = 5$ μm, proving the Cu probe as a better heat sink for its larger thermal conductivity. More insights can be seen from the insets in Fig. 2(a) which show the temperature map of the microcavity system. Clearly, the Cu probe is warmer in a closer distance dissipating heat more efficiently. Indeed, the gap spacing plays a key role to dominant the heating transfer process in such microstructures.

As a comparison, the experimental results in Fig. 2(b) exhibit similar trends for both silica and Cu probes. The average growing rates before $L = 100$ μm are $\eta_{fiber} = 0.21$ MHz/μm and $\eta_{cu} = 0.50$ MHz/μm, respectively. At $L = 5$ μm, the maximum frequency shifts are $\Delta f_{fiber} = 292.26$ MHz and $\Delta f_{cu} = 568.10$ MHz, well matched with the above numerical calculations, i.e., $\Delta f_{fiber-sim} = 316.08$ MHz and $\Delta f_{Cu-sim} = 615.13$ MHz. To further justify this thermal effect, the relationship between input power and $\Delta T$ has been examined with/without external probes. As shown in Fig. 2(c) and (d), the incoming heat causes the temperature rising inside the cavity (from Eq. (2)), causing the shifts of resonances from Eq. (1). As a reference where the fiber probe is far away or absent, the temperature grows linearly with the input power at a rate of $19.69\ \text{mK/μW}$. When the probe is placed at $5\text{μm}$ away, the rate of temperature variation becomes smaller as $17.22\ \text{mK/μW}$ for fiber probe, and $14.15\ \text{mK/μW}$ for cupper probe. The corresponding frequency shift rates of the three lines are $-34.58\ \text{MHz/μW}$, $-30.24\ \text{MHz/μW}$ and $-24.85\ \text{MHz/μW}$, respectively. As a comparison, the measured frequency shift rates in our experiment in Fig. 2(d) are $-33.04\ \text{MHz/μW}$, $-30.40\ \text{MHz/μW}$ and $-26.16\ \text{MHz/μW}$, respectively. These results validate the thermal origin of such fine-tuning mechanism in our microcavity system. As a proof of principle; two representative nonlinear optical applications will be shown below.

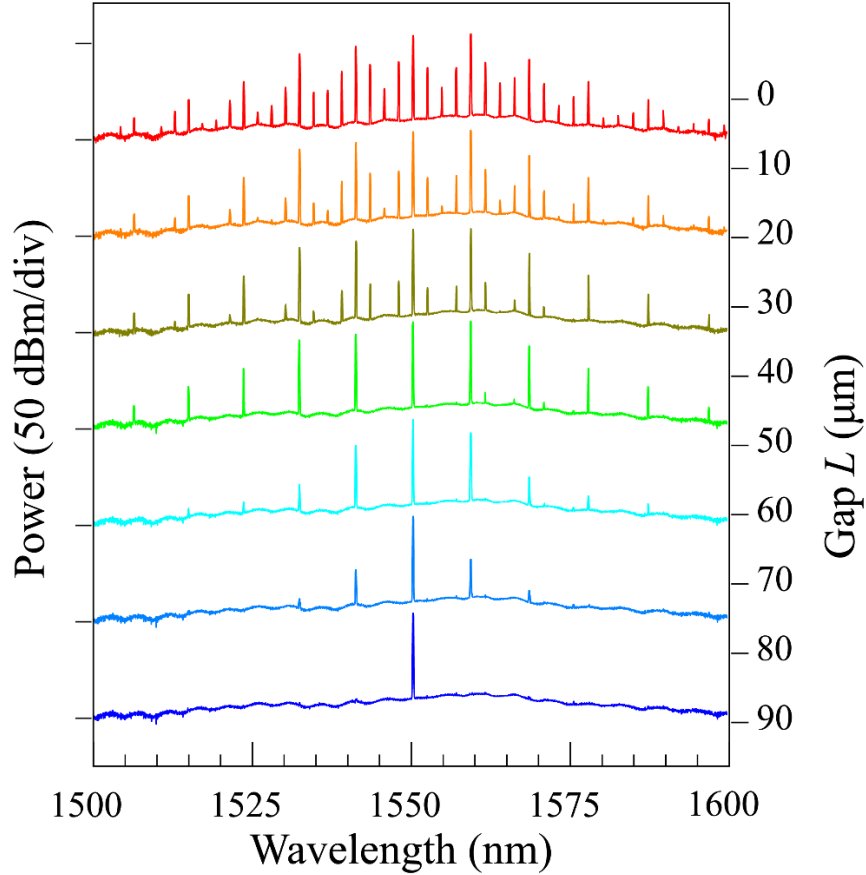

FIG. 3. Fine tuning of optical frequency comb generation. The evolution of frequency comb generation spectrum as the fiber probe approaches to the microsphere. Different OFC families can be chosen by adjusting the gap distance.

We will demonstrate tunable optical frequency comb (OFC) generation and backward stimulated Brillouin scattering (SBS) in the WGM micro-cavity now. Both of the effects are of significance in fundamental science and realistic applications. The feasibility of a WGM based microcavity for fine tuning of frequency comb generation and backward SBS based on such thermal effect is thus demonstrated. In our frequency comb generation, an intense input power (10 mW in our experiment) is injected into the microcavity. The experimental conditions are the same with the aforementioned experiment. The thermal locking is utilized to lock the laser frequency in the WGM resonance. The frequency comb generation in WGM micro-cavities are usually controlled by detuning the pump frequency. It should be noted that the laser frequency (1550.44 nm) is locked to a WGM resonance to generate the frequency comb throughout the tuning process. During the experiment, the fiber tip is firstly placed in the vicinity of the microsphere, such as, with a gap distance of 50 μm. The condition is found by slowly scanning the laser frequency until the frequency comb is generated. Then, the switching of different OFC families is realized by changing the gap distance between the microsphere and the fiber tip. As shown in Fig. 3, the fine tunability of the comb lines is obvious, which proves the feasibility of our scheme.

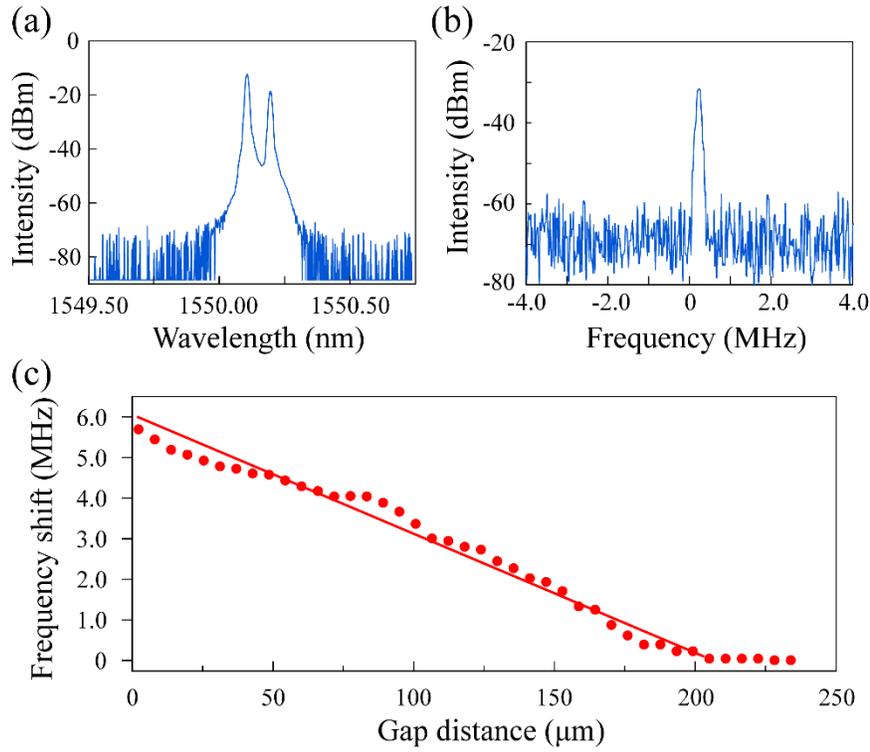

FIG. 4. Tune the backward SBS in WGM microcavity. (a) The optical spectrum of the observed SBS. (b) The electrical spectrum of the observed SBS. The linewidth is about 30 KHz. (c) The central frequency of SBS varies almost linearly while the fiber tip approaches to the microsphere.

We also demonstrate the tunability of SBS frequency in the WGM micro-cavity. Backward SBS phenomena in WGM micro-resonators are the fundamental phenomenon for ultra-narrow linewidth lasers close to quantum limits. The method we proposed may be useful in achieving finely tunable SBS lasers. In the backward SBS experiment, the microsphere diameter is enlarged to approximately 700 μm (by fusing multiple times) to encourage the generation of SBS and also to show the universal applicability of the method. The backward SBS occurs at two optical WGMs separated in frequency exactly by that of the phonon, which has a gain bandwidth of dozens of MHz [28]. They belong to different mode orders (different mode families). The induced frequency shift is therefore slightly different for the two modes. Hence the variation of frequency shift incurs a minute change in the SBS frequency. Fig. 4(a) shows the spectrum of the observed first-order SBS with a pumping power of 5 mW. Careful control is made to ensure higher orders of SBS are not excited. The spectral resolution is 0.01 nm. An SBS frequency of 10.820 GHz is observed when the fiber tip is not introduced, which increases as the fiber tip approaches the microsphere. A tuning span of SBS frequency of about 5.6 MHz is achieved in the experiment as shown in Fig. 4(b). The linewidth at FWHM is approximately 30 kHz. The SBS frequency shift shows a nearly linear dependence on the gap distance as shown in Fig. 4(c). Experimental tests have also been conducted at different resonances, showing different frequency shift span. Spans from 2~10 MHz were the most often observed using microspheres with radius from 250 to 700μm

In conclusion, we have demonstrated a new method for fine tuning of WGM frequency in silica microspheres, using the passive thermal dissipation. Up to 300MHz mode frequency shift are obtained by the fiber probe tip with a MHz resolution. Tuning process is directly related to the gap distance and the materials of the probe tips. With the fine-tuning capability, optical frequency comb generation and backward SBS in the micro-cavities can be precisely tuned. The scheme opens new avenue for micro-cavity applications in nonlinear optics.

**Funding** National Key R&D Program of China (Grant No. 2016YFA0302500, 2017YFA0303700)); Natural Science Foundation of China (Grant No. 11674228, No. 11304201, No. 61475100); National 1000-plan Program (Youth); Shanghai Scientific Innovation Program (Grant No. 14JC1402900); Shanghai Scientific Innovation Program for International Collaboration (Grant No. 15220721400).

**REFERENCES**

[1] Y. Y. Zhi, X. C. Yu, Q. H. Gong, L. Yang, and Y. F. Xiao, Adv. Mater **29**, 1604920 (2017).
[2] Q. F. Yang, X. Yi, K. Y. Yang, and K. J. Vahala, Nature Photon **11**, 560–564 (2017).
[3] A. Imamog, D. D. Awschalom, G. Burkard, D. P. DiVincenzo, D. Loss, M. Sherwin, and A. Small, Phys. Rev. Lett. **83**, 4204 (1999).
[4] M. Aspelmeyer, T. J. Kippenberg, and F. Marquardt, Rev. Mod. Phys. **86**, 1391-1452 (2014).
[5] T. J. Kippenberg, R. Holzwarth, S. A. Diddams, Science **332**, 555-559 (2011).
[6] W. von Klitzing, R. Long, V. S. Ilchenko, Jean Hare, and Valérie Lefèvre-Seguin, "Frequency tuning of the whispering-gallery modes of silica microspheres for cavity quantum electrodynamics and spectroscopy," Opt. Lett. **26**, 166-168 (2001).

[7]A. Dutt, C. Joshi, X. Ji, J. Cardenas, Y. Okawachi, K. Luke, A. L. Gaeta, and M. Lipson, Sci. Adv **4**: e1701858 (2018).

[8]C. Guo, K. Che, Z. Cai, S. Liu, G. Gu, C. Chu, P. Zhang, H. Fu, Z. Luo, and H. Xu, Opt. Lett. **40** (21), 4971-4974 (2015).

[9]B. Wild, R. Ferrini, R. Houdré, M. Mulot, S. Anand, and C. J. M. Smith, Appl. Phys. Lett. **84**, 846 (2004).

[10]K. N. Dinyari, R. J. Barbour, D. A. Golter, and H. Wang, Opt. Express **19**, 17966-17972 (2011).

[11]T. J. Wang, C. H. Chu, and C. Y. Lin, Opt. Lett. **32**, 2777-2779 (2007).

[12]S. Zhu, L. Shi, S. Yuan, X. Xu, and X. Zhang, Opt. Lett. **42**, 5133-5136 (2017).

[13]Q. Ma, T. Rossmann, and Z. Guo, Journal of Physics D: Applied Physics, **41**(24), 245111 (2008).

[14]M. Pöllinger, D. O'Shea, F. Warken, and A. Rauschenbeutel, Phys. Rev. Lett. **103**, 053901(2009).

[15]Q. Xu, S. Manipatruni, B. Schmidt, J. Shakya, and M. Lipson, Opt. Express **15**, 430-436 (2007).

[16]D. Armani, B. Min, A. Martin, and K. J. Vahala, Appl. Phys. Lett. **85**(22), 5439-5441 (2004).

[17]L. Yu, Y. Yin, Y. Shi, D. Dai, and S. He, Optica **3**, 159-166 (2016).

[18]P. Dong, W. Qian, H. Liang, R. Shafiiha, D. Feng, G. Li, J. E. Cunningham, A. V. Krishnamoorthy and M. Asghari, Opt. Express **18**, 20298-20304 (2010).

[19]J. Zhu, S.K. Ozdemir, and L. Yang, Appl. Phys. Lett. 104, 171114 (2014).

[20]K. D. Heylman, N. Thakkar, E. H. Horak, S. C. Quillin, C. Cherqui, K. A. Knapper, D. J. Masiello and R. H. Goldsmith, Nat. Photonics **10**,788–795 (2016).

[21]S. Zhu, L. Shi, B. Xiao, X. Zhang and X. Fan, ACS Photonics, **5** (9), 3794–3800 (2018).

[22]C. H. Dong, L. He, Y. F. Xiao, V. R. Gaddam, S. K. Ozdemir, Z. F. Han, G. C. Guo, and L. Yang, Appl. Phys. Lett. **94**, 231119 (2009).

[23]D. B. Leviton and B. J Frey, Proc. SPIE **6273** 62732 (2006).

[24]See http://accuratus.com/fused.html for "Fused Silica Engineering Properties" (last accessed December 15, 2018).

[25]T. Carmon, L. Yang, and K. J. Vahala, Opt. Express **12**, 4742-4750 (2004).

[26]K. D Heylman and R. H. Goldsmith, Appl. Phys. Lett. **103**, 211116 (2013).

[27]See https://www.comsol.com/products for "COMSOL Multiphysics®" (last accessed December 15, 2018).

[28]R. W. Boyd, *Nonlinear Optics* (Academic Press, Elsevier,2008), P441.